\documentclass[prl,amsmath,amssymb,twocolumn]{revtex4-1}
\usepackage[normalem]{ulem}
\usepackage{amsmath,amssymb}
\usepackage[bookmarks=true,colorlinks,linkcolor=blue,urlcolor=blue,citecolor=blue]{hyperref}
\usepackage{graphicx,amsmath,amsfonts}
\usepackage[usenames]{color}
\usepackage{epstopdf}
\usepackage{verbatim}
\usepackage{amsthm}
\usepackage{relsize}

\newcommand{\be}{\begin{equation}}
\newcommand{\beq}{\begin{eqnarray}}
\newcommand{\eeq}{\end{eqnarray}}

\def \be{\begin{equation}}
\def \ee{\end{equation}}
\def \bw{\begin{widetext}}
\def \ew{\end{widetext}}
\def \ba{\begin{array}}
\def \ea{\end{array}}
\def \bea{\begin{eqnarray}}
\def \eea{\end{eqnarray}}
\def \nn{\nonumber}
\def \half{\frac{1}{2}}

\def \a{{\alpha}}
\def \t{{\theta}}

\def \g{{\gamma}}
\def \D{{\Delta}}
\def \d{{\delta}}

\def \s{{\sigma}}

\def \h{{\eta}}
\def \G{{\Gamma}}

\def \ub{{{\bar u}}}

\def \av#1{{\langle#1\rangle}}

\def \vx{{\bf x}}

\begin{document}
%\title{Superfluidity in two-dimensional driven Bose condensates requires anisotropy}
%\title{Absence of algebraic order in a two-dimensional driven condensate}
%\title{Two-dimensional superfluidity in driven systems requires strong anisotropy}

\title{Two-dimensional superfluidity of exciton-polaritons requires strong anisotropy}

\author{Ehud Altman}
\address{Department of Condensed Matter Physics, Weizmann Institute of  Science,
  Rehovot 76100, Israel}

\author{Lukas M. Sieberer}

\address{Institute for Theoretical Physics, University of Innsbruck, A-6020 Innsbruck, Austria} 
\address{Institute for Quantum Optics and Quantum Information of the Austrian Academy of Sciences, A-6020 Innsbruck, Austria} 

\author{Leiming Chen}
\address{College of Science, The China University of Mining and Technology, Xuzhou Jiangsu, 221116, P.R. China}

\author{Sebastian Diehl}
% 2,3, J. Toner41 Department of Condensed Matter Physics, Weizmann Institute of  Science, Rehovot, Israel2 
\address{ Institute for Theoretical Physics, University of Innsbruck, A-6020 Innsbruck, Austria} 
\address{Institute for Quantum Optics and Quantum Information of the Austrian Academy of Sciences, A-6020 Innsbruck, Austria} 

\author{John Toner}

\address{Department of Physics and Institute of Theoretical Science,  University of Oregon, Eugene OR, 97403, U.S.A.}

%\author{Ehud Altman \\{\small \em Department of Condensed Matter Physics, Weizmann Institute of Science, Rehovot 76100, Israel}}
\begin{abstract}
Fluids of exciton-polaritons, excitations of two dimensional quantum wells in optical cavities, show collective phenomena akin to Bose condensation. However, a fundamental difference from standard condensates stems from the finite life-time of these excitations, which necessitate continuous driving to maintain a steady state. A basic question is whether a two dimensional condensate with long range algebraic correlations can exist under these non-equilibrium conditions. 
Here we show that such driven two-dimensional Bose systems cannot exhibit algebraic
  superfluid order except in low-symmetry, %unless the underlying microscopic system is 
  strongly
  anisotropic  systems. Our result implies, in particular, that recent apparent evidence
  for Bose condensation of exciton-polaritons 
  must be an intermediate scale crossover phenomenon, while the true long
  distance correlations fall off exponentially. We obtain these results through
  a mapping of the long-wavelength condensate dynamics onto the anisotropic
  Kardar-Parisi-Zhang equation.
\end{abstract}
\maketitle

One of the most striking discoveries to emerge from the study of non-equilibrium
systems is that they sometimes exhibit ordered states that are impossible in
their equilibrium counterparts. For example, it has been shown~\cite{TT} that a
two-dimensional ``flock'' - that is, a collection of moving, self-propelled
entities - can develop long-ranged orientational order in the presence of finite
noise (the non-equilibrium analog of temperature), and in the absence of both
rotational symmetry breaking fields and long ranged interactions. In contrast, a
two-dimensional {\it equilibrium} system with short-ranged interactions (e.g., a
two-dimensional ferromagnet) cannot order at finite temperature; this is the
Mermin-Wagner theorem~\cite{Mermin1966}.

In this paper, we report an example of the opposite phenomenon: A {\it driven},
two-dimensional Bose system, such as a gas of polariton excitations in a
two-dimensional isotropic quantum well~\cite{Carusotto2012}, cannot exhibit
off-diagonal algebraic correlations (i.e., two-dimensional superfluidity). In
the polariton gas, the departure from thermal equilibrium is due to the
incoherent pumping needed to counteract the intrinsic losses and maintain a
constant excitation density.

The critical properties of related driven quantum systems have been the subject
of numerous theoretical studies; in certain cases it can be shown that the low
frequency correlation functions induced by driving are identical to those in
equilibrium systems at an effective temperature set by the driving
\cite{Mitra2006,Gopalakrishnan2010,DallaTorre2012,DallaTorre2013,Sieberer2013}. Such emergent  equilibrium behavior %thermalization 
occurs in three dimensional bosonic systems, although non-equilibrium effects  
  {\it can} change the {\it dynamical} critical behavior~\cite{Sieberer2013,Sieberer2013b}. Here, we show that the non-equilibrium conditions imposed by the driving have a much more dramatic effect on \emph{two-dimensional} Bose
  systems: effective equilibrium is never established in the generic
isotropic case; instead, the non-equilibrium nature of the fluctuations
inevitably destroys the condensate at long scales.

This conclusion follows from the known ~\cite{Shivashinsky1977,Kuramoto1984,Grinstein1993,Grinstein1996}  %an exact mapping, {\cred long known\cite{Grinstein} in the context of 
connection between the Complex Ginzburg-Landau equation (CGL) (which describes
the long wavelength dynamics of a driven
condensate) and the Kardar-Parisi-Zhang (KPZ) equation~\cite{Kardar1986}, or, in the anisotropic case, the anisotropic KPZ equation~\cite{Wolf1991}, which were 
originally formulated to describe randomly growing interfaces. The
non-equilibrium fluctuations generated by the drive translate into the
non-linear terms of the KPZ equation. 

Our results suggest that recent
experiments~\cite{Kasprzak2006,Balili2007,Deng2007, Roumpos2012} done with
isotropic semiconductor quantum wells purporting to show evidence for the long
sought~\cite{Imamoglu1996} Bose condensation of polariton excitations are in
fact observing an intermediate length scale crossover phenomenon, and not the
true long-distance behavior of correlations. We remark that earlier work, which predicted long range algebraic order in two-dimensional driven condensates~\cite{Chiocchetta2013},  relied on a linear (Bogoliubov) theory, which  may appear on intermediate scales but,  as  our analysis shows, is invalidated at long distances due to the relevant non-linearity. 

On the other hand,  performing the  same mapping on the {\it anisotropic} CGL leads, as  noted by Grinstein et al. \cite{Grinstein1993,Grinstein1996}, to the {\it anisotropic}  KPZ equation. This suggests,
as  also noted by those authors, that algebraic order can
prevail if the system is anisotropic. Then
the transition to the disordered phase occurs by a standard equilibrium-like
Kosterlitz-Thouless transition. This requires very strong anisotropy, which 
may seem unnatural in the case of exciton-polaritons in two dimensional quantum wells.
However, mapping a realistic model 
of such a system to the anisotropic KPZ equation shows that the anisotropy of the KPZ non-linearities is a function of the driving laser power. Surprisingly, we find that even if the intrinsic anisotropy of the system is moderate, the effective anisotropy increases with pump power and eventually passes the threshold allowing for  an effective equilibrium description. Then, not only does an algebraically ordered phase occur, but it does so in a {\it reentrant} manner: the phase is  entered, and then left, as the driving laser power is increased.

% Our results also apply to a far broader class of systems; namely, any system described by the noisy complex Ginzburg-Landau equation~\cite{Graham1990,Aranson2002}. This includes, in addition to the driven BEC problem that motivated us,  nonlinear chemical waves~\cite{Kuramoto1984}, driven  superconductors~\cite{Mitra2006}, driven noisy oscillators modelling the physics of hearing~\cite{julicher05,julicher10},
%and a variety of pattern-forming dynamical systems~\cite{Cross1993}.{\cblue While this connection is already well-known for isotropic systems, the connection in the anisotropic cases leads to the new prediction that anisotropic examples of the above systems can have defect-free phases.}}

\noindent {\em Model --} 
%The simplest model for Bose-Einstein condensation has a complex scalar order parameter field $\psi$, which can be thought of as the macroscopic wavefunction of the condensate. Since we are interested in a driven, non-equilibrium system, we cannot proceed as in equilibrium, i.e., by considering the partition function associated with a ``Landau-Ginzburg-Wilson free energy'' that is a functional of $\psi$. Rather, we must formulate an equation of motion consistent with the symmetries of the systems. Here, the symmetries are translation invariance and the usual invariance under uniform changes in the phase of the wavefunction: $\psi\rightarrow\psi e^{i\phi}$, where $\phi$ is an arbitrary constant. In order for the latter to remain a symmetry for short-lived excitations such as polaritons, the driving field which maintains them at a constant density must be incoherent, so that it does not imprint a phase on the polaritons it excites.
The dynamics of a driven-dissipative system like a polariton condensate, described by a complex scalar order parameter field $\psi$, is
determined both by coherent processes, such as the dispersion and scattering
between polaritons, 
and independent dissipative processes induced by loss and the pumping field.  A model
of the condensate dynamics that incorporates these processes is %and respects the aforementioned symmetries is \cite{Sieberer2013}:
\be
\partial_t\psi(\vx,t) = - {\d H_d\over \d \psi^*}+i{\d H_c\over \d\psi^*} +\zeta({\bf x},t).
\label{EOM}
\ee
Here, the effective Hamiltonians $H_\ell$ ($\ell=c,d$) that generate the coherent and dissipative dynamics respectively read
%\bw
\be
H_\ell= \int_{x,y} \left[r_\ell |\psi|^2+K^x_\ell |\partial_x\psi|^2+ K^y_\ell |\partial_y\psi|^2+\half u_\ell |\psi|^4\right]~.
\label{Hcd}
\ee
%\ew
The last term $\zeta(\vx,t)$ in Eq.~(\ref{EOM}) is a zero mean Gaussian white noise with short-ranged spatiotemporal correlations: $\av{\zeta^*(\vx,t)\zeta(\vx',t')}=2\sigma\d^d(\vx-\vx')\d(t-t')~, \av{\zeta(\vx,t)\zeta(\vx',t')}=0$.

Eq.~(\ref{EOM}) is widely known as the complex Ginzburg-Landau
equation~\cite{Cross1993,Aranson2002}, or in the context of polariton
condensates, as the dissipative Gross-Pitaevskii
equation~\cite{Wouters2007,Keeling2008}, although usually only the isotropic
(i.e., $K^x_\ell=K^y_\ell$), noise free ($\zeta=0$) case is considered (but see
[\onlinecite{Graham1990}]).  Modifications of this equation, e.g., including
higher powers of $\psi$ and $\zeta$, higher derivatives, or combinations of the
two, can readily be shown to be irrelevant in the Renormalization Group (RG)
sense: they have no effect on the long-distance, long-time scaling
properties of either the ordered phase, or the transition into it \footnote{The
  gradient terms shown are the only ones at second order in gradients allowed in
  systems which have inversion symmetry about either the $x$ or the $y$ axis; we
  will restrict our discussion here to such systems.}.

Each of the parameters appearing in the model has a clear physical origin, as we now review. The coefficient $r_d$ is the single particle loss rate  
$\gamma_l$ (spontaneous  decay) offset by the pump rate $\gamma_p$, that is,
$r_d=\gamma_l-\gamma_p$.
In contrast, $r_c$ is an effective chemical potential, which 
is {\em completely arbitrary}. Indeed, it can be adjusted by a temporally local gauge transformation $\psi(\vx,t)=\psi'(\vx,t)e^{i\omega t}$, such that $r_c'=r_c-\omega$. In the following, we choose $r_c$  so that, in the absence of noise, the equation of motion has a stationary, spatially uniform solution.
%By  changing  variables $\psi(\vx,t)=\psi'(\vx,t)e^{i\omega t}$ we can obtain  Eq.~(\ref{EOM}) for $\psi'(\vx,t)$, with only $r_c$ changed, by $r_c\rightarrow r_c-\omega$. Thus, by a suitable choice of $\omega$, we can make $r_c$ take on any value we like. This is very similar to a ``gauge choice''. We will henceforth choose $r_c'$ so that,  in the absence of noise, the equation of motion has a stationary, spatially uniform solution.  We will also  drop the prime in the following.

The term proportional to $u_c$ is the pseudo-potential which describes the elastic scattering of two polaritons, whereas $u_d$ is the non-linear loss  or, alternatively, a reduction of the pump rate with density,  that ensures saturation of particle number. 
The coefficients $K^{x,y}_c= \hbar^2/(2m_{x,y})$, where $m_{x,y}$ are  the eigenvalues of the effective polariton mass tensor, with principal axes ${x,y}$. Under typical circumstances, the diffusion-like term $K_d$ is expected to be small, but is 
 allowed by symmetry, and so will always be generated~\cite{Wouters2010,Wouters2010a}.   Finally, the noise is given by the total rate of particles entering and leaving the system.  In polariton condensates, where $u_d$ reflects a non-linear reduction of the pumping rate rather than an additional loss mechanism (see supplementary information), the noise level at steady state is simply set by the single particle loss, i.e.  $2\sigma=2\gamma_l$.~ \footnote{In reality the system might be coupled also to a particle conserving bath, such as phonons in the solid, which we have not included. While such a coupling is irelevant in the RG sense, if it is strong and induces fast relaxation toward the bath equilibrium, it may renormalize the parameters of the stochastic equation of motion (\ref{EOM}). However, such a bath will not restore detailed balance, and none of the universal results presented here will change.}

Before proceeding, it is important to clarify under what conditions
Eq.~(\ref{EOM}) describes an effective thermal equilibrium at {\it all}
wavelengths. Imposing the additional condition that the field follows a thermal Gibbs distribution at steady state  translates to the simple requirement $H_d=R H_c$, where
$R$ is a multiplicative constant~\cite{graham73,deker75,Graham1990,tauberreview}. This condition can also be seen as a symmetry of the dynamics, which ensures detailed balance~\cite{Sieberer2013,Sieberer2013b}. 
%If the steady state of such a stochastic equation is to be described by a Gibbs distribution of the field, it should satisfy a specific relation between the Hamiltonian component of the evolution generated by Poisson-brackets and the dissipative component~\cite{graham73,deker75,Graham1990,tauberreview}.  Applied to the problem at hand, with Poisson brackets $\{\psi(\mathbf{x}),\psi(\mathbf{x}')^*\}=\d(\mathbf{x}-\mathbf{x}')$, the equilibrium condition translates to the simple requirement $H_d=R H_c$, where 5$R$ is a multiplicative constant.
In a driven system, the relation  $H_d=R H_c$ is not satisfied in general, because the dissipative and coherent parts of the
dynamics are generated by independent processes.  This relation {\it can}, however,  arise as an emergent symmetry at low frequencies and long
wavelengths. This was shown to be the case for a three-dimensional driven
condensate~\cite{Sieberer2013,Sieberer2013b}. Below we shall derive the hydrodynamic
long-wavelength description of a two-dimensional driven condensate and determine
if it flows to effective  thermal equilibrium.

\noindent {\em Mapping to a KPZ equation --} In the long-wavelength limit, Eq.~\eqref{EOM} reduces to a KPZ
equation~\cite{Kardar1986} for the phase variable~\cite{Grinstein1993}.  As in equilibrium, in a hydrodynamic description of the condensate  the order-parameter field is written
in the amplitude-phase representation as
$\psi(\vx,t)=(M_0+\chi(\vx,t))e^{i\t(\vx,t)}$. Integrating out the gapped amplitude mode, and keeping only terms which are not irrelevant in the sense of the renormalization group, we obtain a closed equation for $\t$ (see Method section for details)
\be
\partial_t\t= D_x\partial_x^2\t+D_y\partial_y^2\t+{\lambda_x\over 2}(\partial_x\t)^2+{\lambda_y\over 2}(\partial_y\t)^2+{\bar\zeta}({\bf x},t)~,
\label{akpz}
\ee
with  ($\alpha = x,y$):
\bea
D_{\alpha}&=&K^{\alpha}_d\left[1+{K^{\alpha}_c \over  K^{\alpha}_d } {u_c\over   u_d}\right]~,
\label{D}\\\nonumber
\lambda_{\alpha}&=&2 K^{\alpha}_c\left[{K^{\alpha}_d\over  K^{\alpha}_c}{u_c\over u_d }-1\right] 
~.\nn
\label{lambda}
\eea
and noise level (replacing $\sigma$ in the noise correlations above)
%{\cmag
\bea
\D&=&%{\sigma(1+u_c^2/u_d^2)\over  2M_0^2}=(1+u_c^2/u_d^2){\g_p
%+\g_l+u_d M_0^2\over 4M_0^2}\nn\\ &=&
 {(u_d^2+u_c^2)\g_l\over  2u_d(\g_p-\g_l)}~.
\label{kpznoise}
\eea
%[I changed the above expression to be the case relevant for polariton condensates where the only true loss term is $\gamma_l$ and therefore $\s=\gamma_l$].}
Eq.~(\ref{akpz}) is the anisotropic KPZ equation, originally formulated to describe 
the roughness of a growing surface due to random deposition of particles on it~\cite{Kardar1986, 
Wolf1991}, in which case $\t$  is the height of the interface. It reduces to the {\it isotropic} KPZ equation when $D_x=D_y$ and $\lambda_x=
\lambda_y$.   This reduction can also be achieved by
 a trivial rescaling of lengths
if $\G\equiv{\lambda_y D_x\over \lambda_x D_y}=1$. Thus, when $\G \neq 1$, the system is anisotropic.  

Crucially, the presence of the non-linearity directly reflects non-equilibrium conditions \footnote{ More precisely, the KPZ equation describes non-equilibrium conditions in $d>1$. In fact, in one dimension, it can be mapped ``accidentally'' onto the noisy Burgers equation, and thus may occur also in an equilibrium context. An explicit example has been identified in \cite{Lamacraft2013}.}. Indeed, the coefficients $\lambda_x,\lambda_y$ that measure the deviation from thermal equilibrium vanish identically when the conditions 
\bea
K^x_c/K^x_d=K^y_c/K^y_d=u_c/u_d~, 
\eea 
which follow from the equilibrium requirement that  $H_d=R H_c$, are met. 

It is furthermore important to note that our KPZ model differs from that formulated for a description of randomly growing interfaces~\cite{Kardar1986} in that the analog of the interface height variable in our model is actually a compact phase; hence, topological defects in this field are possible.  This difference with the conventional KPZ equation also arises  in ``Active Smectics'' [\onlinecite{Chen2013}]. 

Analysis of Eq.~(\ref{akpz}) in absence of vortices is the analogue of the low temperature spin-wave (linear phase fluctuation) theory of the equilibrium $XY$ model. %The crucial difference here is the appearance of the nonlinear terms with coefficients $\lambda_x,\lambda_y$, which are a direct measure of the deviation from thermal equilibrium. They vanish identically when the equilibrium conditions 
%\bea
%K^x_c/K^x_d=K^y_c/K^y_d=u_c/u_d~, 
%\eea 
%which follow from the equilibrium requirement that  $H_d=R H_c$, are met. 
Indeed, without the non-linear terms,  the KPZ equation reduces to linear diffusion, which would bring the field to an effective thermal equilibrium with power-law off-diagonal correlations (in $d=2$). A transition to the disordered phase in this equilibrium situation can occur only as a Kosterlitz-Thouless (KT) transition through proliferation of topological defects in the phase field.

\begin{figure}
 \includegraphics[width=0.5\textwidth]{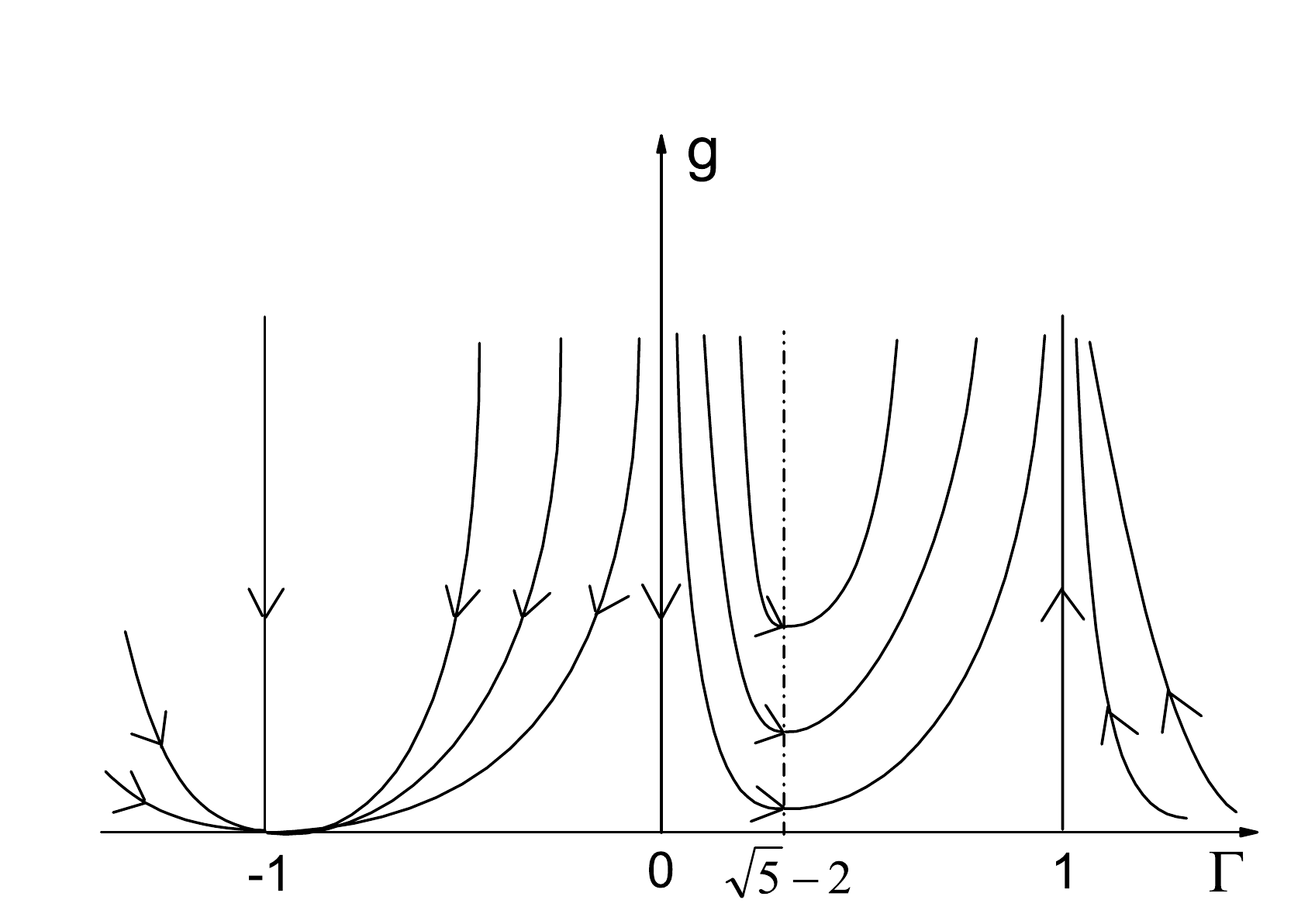}
 \caption{\label{fig:FGraph}The RG flow in the $\Gamma$-$g$ parameter space for anisotropic driven BEC in $d=2$. For $\Gamma<0$ and $g>0$, all flow lines go  to a stable fixed point $(-1,0)$; for $\Gamma>0$ and $g>0$, all flow lines go to infinity,  and approach the isotropic limit $\Gamma=1$.}
\end{figure}

In a driven condensate, the non-linear terms are in general present, and in two
dimensions have the same canonical scaling dimension as the linear terms. A more
careful RG analysis is therefore required to determine how the system behaves at
long scales even without defect proliferation. Such an analysis has been done in
Refs.~\cite{Wolf1991} and~\cite{Chen2013} for the anisotropic KPZ equation. In
this case, the flow is closed in the two parameter space of scaled non-linearity
$g\equiv{\lambda_x^2\D\over D_x^2\sqrt{D_x D_y}}$ and scaled anisotropy
$\G\equiv{\lambda_y D_x\over \lambda_x D_y}$, and is given, to leading order in
$g$, by: \bea
{d g\over d l}&=& { g^2\over 32\pi}(\G^2+4\G-1),\nn\\
{d\G\over d l}&=&{\G g\over 32\pi}(1-\G^2)~.
\label{recrels}
\eea
These flows are illustrated in Fig.~\ref{fig:FGraph}.
We see that in an isotropic system, $\Gamma=1$,  and  the nonlinear coupling $g$, which embodies the non-equilibrium fluctuations, is relevant. Moreover, for a wide range of anisotropies (namely, all $\G>0$) the flow is attracted to the isotropic line: the system flows to strong coupling,
with emergent rotational symmetry. On the other hand, if the anisotropy is sufficiently strong, so that $\G<0$, the non-linearity becomes irrelevant and the system can flow to an effective equilibrium state at long scales.

We will now  discuss the physics of these two 
regimes, starting with  the isotropic case, which is most relevant to current experiments with polariton condensates.

\noindent {\em Isotropic systems --} As noted above, rotational symmetry is emergent at long scales if the anisotropy is not too strong at the outset.  This is also the  regime in which current experimental quantum well polaritons lie.  We therefore consider this case first.  

On the line $\G=1$, the scaling of the non-linear coupling $d g/dl=g^2/ 8\pi$ drives $g\to\infty$; in the growing surface problem the system goes to the ``rough'' state, with height fluctuations scaling algebraically with length. The analogous behavior in the phase field $\theta$ would lead  
 to
 stretched exponentially decaying order parameter correlations. However, the fact that the phase field is compact implies that topological defects (vortices) in this field exist. Our expectation, based  on analogy with equilibrium physics (which admittedly may be an untrustworthy analogy), is that vortices will unbind at the strong coupling fixed point of the KPZ equation. If this happens, it  will lead to simple exponential correlations. Testing  this expectation 
will be the object of future work. %mention phase vs. defect turbulence here seen in numerical work?}
\begin{figure}[t]
  \includegraphics[width=1\columnwidth]{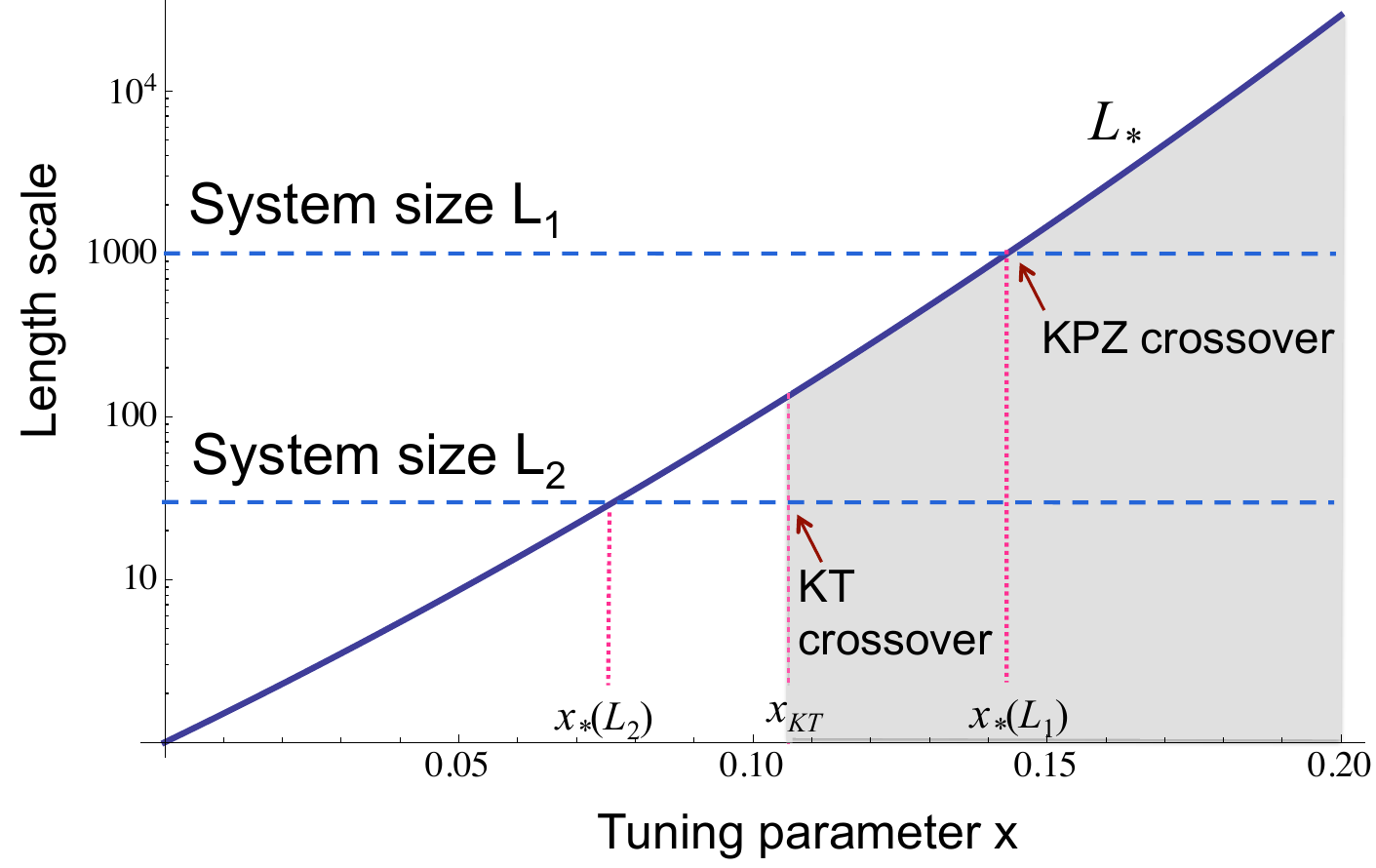}
  \caption{Dependence of the emergent KPZ length scale $L_*$ (in units of the microscopic healing length) on the tuning parameter $x=\g_p/\g_l-1$. This curve was obtained by inserting the expression Eq.~(\ref{g_02band}) for the bare coupling $g_0$ into Eq.~(\ref{L_*}) for $L_*$. While $L_*$ is exponentially large when $x>{\bar u}{\bar\g}^2/2\pi$, it goes to a microscopic value $\xi_0$ at the mean field transition $x=0^+$. The shaded region marks the scales at which a system would exhibit algebraic correlations.  Upon decreasing the tuning parameter $x$, a finite system will lose its algebraic order in one of two ways: (1) When $L_*$ falls below system size, as in the case of system $L_1$ shown, or (2) in a KT transition before $L_*$ falls below system size, as in the case of system $L_2$. Here we have used $\bar\g=0.7$, $\bar u=0.5$.}
\label{fig:lengths1}
\end{figure}

We have  thus established that the non-linearity, no matter how weak, destroys the condensate at long distances, leading to either stretched or simple exponential decay of correlations  throughout the isotropic regime.
However, the effects of the nonlinearity only become apparent when $g$ gets to be of order one. Solving the scaling equation we see that this occurs at the characteristic ``RG time'' $l_*=8\pi/g_0$;   the corresponding  length scale is:
\be
L_{*}=\xi_0 e^{\ell^*}=\xi_0e^{8\pi/g_0}~,
\label{L_*}
\ee
where $\xi_0$ is the mean field healing length of the condensate.
If  the bare value of $g_0$ is small, then the scale $L_{*}$ can be huge. On length scales smaller than $L_{*}$,  the system is governed by the linearized isotropic KPZ equation, which, as noted earlier, is the same as an equilibrium $XY$ model. Thus, all of the equilibrium physics associated with two-dimensional BEC, including power law correlations and a Kosterlitz-Thouless defect unbinding transition, can appear in a sufficiently small system. 

As parameters, such as the pump power, are changed, the system can lose its apparent algebraic order in one of two ways: (i) the KPZ length $L_*$ is gradually reduced below  the system size, or (ii) $L_*$ remains large while the correlations within the system size $L$ are destroyed by unbinding of vortex anti-vortex pairs at the scale $L$. The latter type of crossover would appear as a KT transition broadened by the finite size. Of course, for any given set of system parameters, a sufficiently large system ($L>L_*$)  will always be disordered.

We shall now discuss how the system parameters determine what type of crossover,
if any, will be seen in an experiment. We assume that the main tuning parameter
is the pump power $\g_p$, and it will be convenient to track the behavior as a
function of a dimensionless tuning parameter $x=\g_p/\g_l-1$, and set $K_d=0$ since this parameter is thought to be small in current experimental realizations.  %But other parameters can also be tuned continuously in an experiment; for example, the effective polariton mass can be varied by  illuminating different regions in a spatially varying quantum well. 
In the supplementary material, we derive the parameters of the KPZ equation for a
realistic model of a polariton condensate, described by a two fluid model which
includes an upper polariton band, into which polaritons are pumped, and a lower
band, where the condensate forms. In particular, we 
 obtain an expression for the bare dimensionless coupling constant $g_0$ in this model,  which measures the bare deviation from equilibrium:
\be
g_0={\D\lambda^2\over D^3}=2{ \bar{u} \bar{\g}^2}\,\left({\bar{\g}^2+(1+x)^2 \over x (1+x)^3}\right).
\label{g_02band}
\ee
Here ${\bar u}\equiv u_c/K_c$ is the dimensionless interaction constant, and $\bar \gamma\sim \gamma_l$  the dimensionless loss rate (see supplementary material).  Note that $g_0$ diverges as we approach the mean field transition at $x\to0^+$, while it decays as $1/x^2$ as $x\to \infty$ at very high pump power.

Hence, in the latter regime the KPZ length scale $L_*=\xi_0\exp(8\pi/ g_0)$ is certainly much larger than any reasonable system size. As the pump power is decreased, and the system approaches the mean field transition at $x=0$,  $L_*$ drops sharply to a microscopic healing length $L^*\approx \xi_0$.  $L_*$ drops below the system size when $x\lesssim x_*$, where
\be
{x_*(1+x_*)^3\over {\bar\g}^2+(1+x_*)^2}\approx {{\bar u}{\bar\g}^2\over 4\pi}\ln(L/\xi_0)~.
\label{xs}
\ee
For pump powers corresponding to $x>x_*$, the system will appear to be  at effective equilibrium,  and, hence,  may sustain power law order within its confines, whereas for pump power $x<x_*$, the non-equilibrium fluctuations become effective and destroy the algebraic correlations at the scale of the system size. However, it is possible that this crossover at $x_*$ is preceded by unbinding of vortices at values of $x=x_{KT}>x_*$, while the finite system is still at effective equilibrium.

To  determine which crossover occurs in a particular system,  let us estimate the value of the tuning parameter $x_{KT}$ at which the putative Kosterlitz-Thouless transition {\it would} occur if the non-linear term $\lambda$ vanished, or was negligible. 
Then Eq.~(\ref{akpz}) obeys a fluctuation-dissipation relation with a temperature set by the noise $T=\D$. The KT transition would occur for an equilibrium $XY$ model approximately at the point where $\D/D=\pi$. Expressing both $\D$ and $D$  in terms of the tuning parameter $x$, we obtain the equation for the critical value $x_{KT}$ at which the Kosterlitz-Thouless transition will appear to occur:
\be
{x_{KT}(1+x_{KT})\over {\bar\g}^2+(1+x_{KT})^2}\approx {{\bar u}\over 2\pi}.
\label{xkt}
\ee

We can solve the equations for $x_*$ and $x_{KT}$ in certain simple limits. For
example, if we assume weak interactions ${\bar u}\ll 2\pi$, and in addition
$({\bar u} /4\pi){\bar\g}^2\ln (L/\xi_0)\ll 1$, then $x_*$ and $x_{KT}$ are
given approximately by $x_*\approx ({\bar u} /4\pi){\bar\g}^2(1+ {\bar\g}^2)\ln
(L/\xi_0)$, and $x_{KT}\approx ({\bar u} /4\pi)(1+{\bar\g}^2)$. Under these
conditions we expect to see a crossover controlled by vortex unbinding through
the KT mechanism, i.e., $x_{KT} >x_*$, if the system size is $L<\xi_0
\exp(2/{\bar\g}^2)$. For larger system size the crossover will be controlled by
the nonlinearities of the KPZ equation. This crossover behavior is summarized in
Fig. \ref{fig:lengths1}.

\noindent {\em Strong anisotropy --}  If the bare value of the anisotropy parameter is negative $\G<0$, then the RG equations (\ref{recrels}) lead to a fixed point at $g=0$.
Because the non-linear $\lambda_{x,y}$  terms in (\ref{akpz})  are irrelevant in this region of parameter space, the linear (and, hence, equilibrium) version of the theory applies. Hence it is possible, for
$\G < 0$, to obtain both a power law phase and a KT defect unbinding transition out of it.

To estimate the extent of this phase, we can utilize the RG flow of the anisotropic KPZ equation for $\G<0$ analyzed in Ref. [\onlinecite{Chen2013}]. In principle, we should add to these recursion relations terms coming from the vortices. Instead, we will follow reference~\cite{Chen2013} and assume that the vortex density is low enough that vortices only become important on length scales far longer than those at which the nonlinear effects have become unimportant (i.e., those at which the scaled non-linearity $g$ has flowed to nearly zero). If this is the case, then we can use the recursion relations Eq.~(\ref{recrels}) for our problem, despite the fact that they were derived neglecting vortices. 

Our strategy is then to use those recursion relations to flow to the linear regime, which, as noted earlier, is equivalent to an equilibrium $XY$ model. 
In this regime, vortex unbinding is controlled by the (bare) parameters $\kappa_0\equiv\Delta/\sqrt{D_{x}D_y}$  (cf.\ Eq.~\eqref{kpznoise}, or 
\eqref{Delta} in the supplementary material for the two-band model),  giving the scaled noise level and replacing the temperature of the equilibrium problem as above, as well as the scaled anisotropy $\G_0$. Following~\cite{Chen2013}, %The critical point for vortex unbinding can be estimated by solving for the renormalized  scaled noise $\kappa(\ell\to\infty)\equiv\kappa(\infty)$ as a function of the bare value using the RG equations of the non-compact KPZ equation; this involves additional recursion relations for $D_\alpha$ and $\D$ as well as Eq.~(\ref{recrels}); for details see reference~\cite{Chen2013}. This gives $\kappa(\infty)=-\kappa_0(1-\G_0)^2/(4\G_0)$. The KT transition occurs at the point where this renormalized value $\kappa(\infty)$ of $\kappa$ reaches $\pi$. Hence 
the  phase boundary in the $\kappa_0$-$\G_0$ plane is then a locus %given by the curve
in the plane of bare scaled noise and anisotropy parameters given by ~\cite{Chen2013}
\begin{eqnarray}
\kappa_0=-{4\pi\Gamma_0\over\left(1-\Gamma_0\right)^2}~,
\label{phaseboundary}
\end{eqnarray}
see the supplementary materials for more details.
%The assumption in deriving this curve was that the dominant contribution to the stiffness renormalization comes from the non-linear fluctuations,  rather than from bound vortex-antivortex pairs, which have been neglected. 

\begin{figure}
\includegraphics[width=0.5\textwidth]{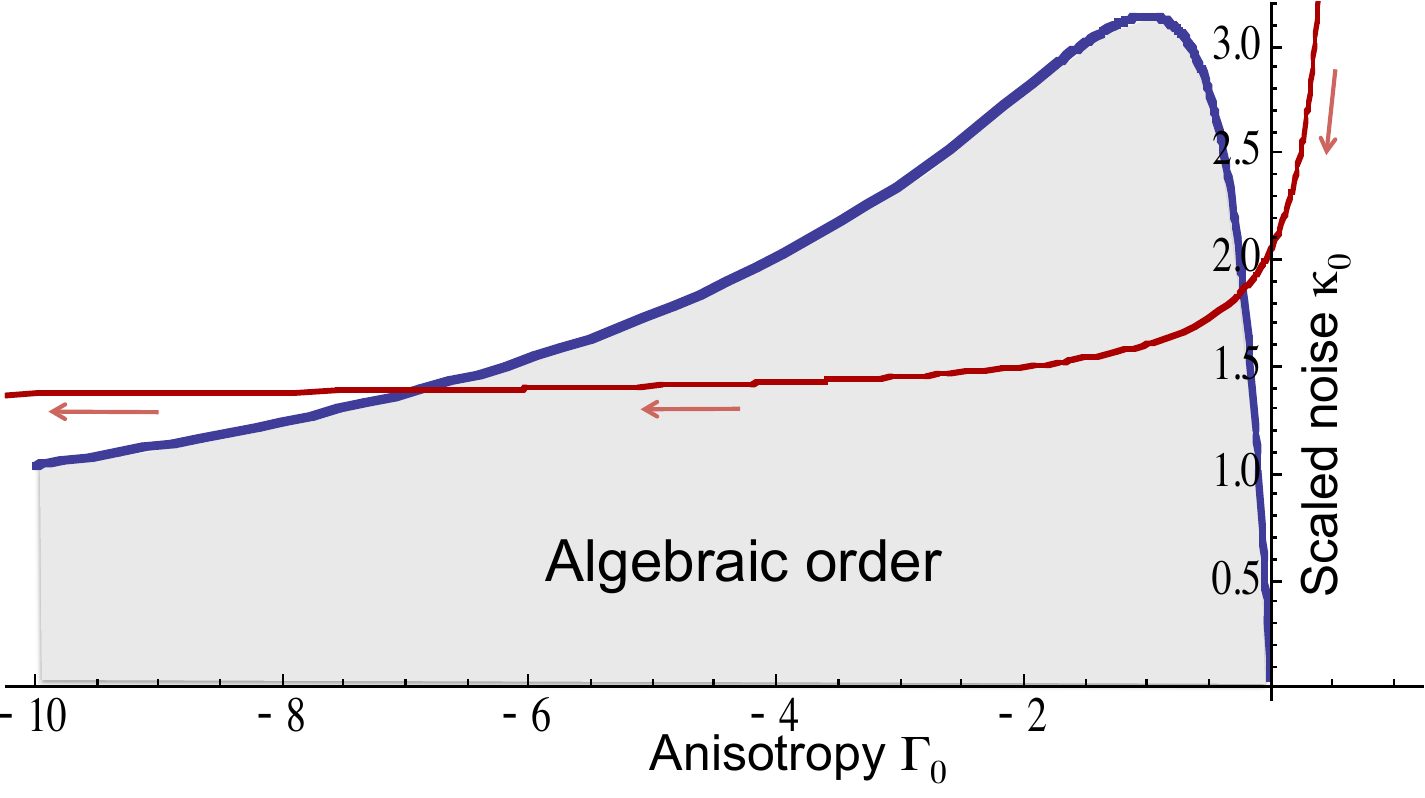}
 \caption{\label{fig:anisotropic} Phase diagram of a generic  anisotropic system exhibiting reentrance. The thin line marks the natural trajectory in an experiment in which only the pump-power is varied. The arrows mark the direction of increasing pump power. Such an experiment will see a reentrant behavior, where the system starts in a disordered state, enters the power-law superfluid and then goes back to a disordered state. Here we have used the two band model of the supplementary material, with $\bar\g=1/2$, $\ub=2$, $\nu_x=1/8$ and $\nu_y=1/4$.}
\end{figure}

There is a broad range of parameters for which a 
system enters a regime 
$\Gamma_0= (D_x\lambda_y)/(D_y\lambda_x)<0$, in which true power-law order and a KT transition exist. Within the ``two-band polariton model'' (see supplementary material), as a function of microscopic parameters we obtain 
%A natural and simple model of such a system is the ``two-band polariton model'' discussed earlier, and treated in the supplemental material. Two important parameters determining the behavior of the anisotropic system are the ratios of the dissipative to coherent phase stiffnesses along the two directions, $\nu_{\a}=K_d^\a/K_c^\a$. Using the expressions for $D_\alpha$, $\lambda_\alpha$, and $\D$ in the two-band model we can  easily obtain the dependence of the  bare anisotropy parameter $\G_0$ and the  bare scaled noise $\kappa_0$ on the tuning parameter $x$:
\bea
\G_0&=&{\left[\nu_y(1+x)-{\bar \gamma}\right]\left[\nu_x{\bar \gamma}+1+x\right]\over\left[\nu_x(1+x)-{\bar \gamma}\right]\left[\nu_y{\bar \gamma}+1+x\right]}
\label{G_0},\\
\kappa_0&=&{\bar{u}\over 2x}\,{[\bar{\gamma}^2+(1+x)^2]\over\sqrt{[\nu_y\bar{\g}+(1+x)][\nu_x\bar{\g}+(1+x)]}},
\label{kappa_0}
\eea
with the ratios of the dissipative to coherent phase stiffnesses along the two directions, $\nu_{\a}=K_d^\a/K_c^\a$. Now consider gradually increasing the pump power, and hence $x$, from the mean field threshold $x=0$.
For system parameters $\bar\gamma >\nu_y>\nu_x$, $\G_0(x)$ starts out positive at
$x=0$, is reduced to negative values as $x$ is increased past  $x={\bar\gamma\over\nu_y}-1$, and eventually runs off to $\G_0=-\infty$ at a finite value of $x$ (namely, $x={\bar\gamma\over\nu_x}-1$). If at the same time $\ub$ is sufficiently small, then the experimental trajectory in the $\kappa_0-\Gamma_0$ plane is guaranteed to cross the dome marking the condensate (algebraic order) phase as determined in Eq.~(\ref{phaseboundary}).  
The condition on $\bar u$ for this crossing to occur is
\be
\ub<2\pi{\left(\bar\gamma-\nu_y\right)\over \bar\gamma\nu_y(1+\nu_y^2)}~.
\label{ubound}
\ee

Thus, we not only naturally achieve the ordered phase in this anisotropic system by varying the driving, but we do so in a {\it reentrant} manner: we enter the phase, and then leave it again, as the driving is increased.
The analysis for $\gamma>\nu_y>\nu_x$ is the same if we take $\G\to 1/\G$.

\noindent{\em Outlook --} %{\color{blue} I think they do not like conclusions which just reproduce the text, they will maybe bar it out.} We have shown that quasi-long-ranged polaritonic Bose Einstein Condensation cannot exist in a two-dimensional system, unless the system is strongly anisotropic.  This result is obtained by mapping  of the condensate dynamics to the (compact) anisotropic Kardar-Parisi-Zhang equation.  
%Non-equilibrium fluctuations generated by the drive translate to the non-linear term in the KPZ equation, which disorders  the condensate at long scales. We remark that earlier work, which predicted long range algebraic order in two-dimensional driven condensates~\cite{Chiocchetta2013},  relied on a linear (Bogoliubov) theory, which  may appear on intermediate scales but,  as  our analysis shows, is invalidated at long distances due to the relevant non-linear term. 
%
%It is worth noting that recent analysis of the dissipative Ginzburg-Landau equation without noise found that the mean field condensate is destroyed, through a completely different mechanism, by static disorder~\cite{Janot2013}. The consequences of the interplay of noise and the static disorder is an interesting question for future investigations.
%
%We have also shown that for sufficiently  strong breaking of four- or six- fold rotational symmetry,  the non-linearity becomes irrelevant, allowing the establishment of effective equilibrium. In this case, the system can develop algebraic correlations through a standard Kosterlitz-Thouless transition. 
Our analysis can be extended to three dimensions. There, for weak deviations from equilibrium, i.e. a small bare value of the KPZ non-linearity, it predicts a true Bose condensate which may be established through the dynamical phase transition described in~\cite{Sieberer2013}. However, beyond a critical strength of the equilibrium deviation, one may also encounter a different, non-equilibrium transition controlled by a strong coupling fixed point of the three dimensional KPZ equation~\cite{Fisher1992}. This opens up the possibility for a new non-equilibrium phase of matter with short-ranged order, distinct form the usual uncondensed state in that vortex loops do not proliferate. This will be explored in future work.

\section{Methods}

Here we review the mapping~\cite{Grinstein1993}, in the long-wavelength limit, between  the
model~(\ref{EOM})  and an anisotropic KPZ
equation~\cite{Kardar1986, Wolf1991}.  We work 
in the amplitude-phase representation
$\psi(\vx,t)=(M_0+\chi(\vx,t))e^{i\t(\vx,t)}$, with $M_0$, $\chi$, and $\theta$
all real. Here $M_0$ is determined by requiring that $\chi=0$, $\theta=0$ is a
static uniform solution of Eq.~(\ref{EOM}) in the absence of fluctuations
($\zeta(\vx,t)=0$). The real and imaginary parts of Eq.~(\ref{EOM}) then
give $M_0^2=-r_d/u_d$ and $ r_c=-u_c M_0^2$, respectively.  We can satisfy the
second condition by exploiting the freedom to choose $r_c$ mentioned in the main text.  As
explained there, by varying the strength of the pump laser, one can
experimentally control $r_d$, which determines the amplitude $M_0$. The mean
field transition occurs at the point $r_d=0$ (i.e., when $\g_p=\g_l$), where the
amplitude $M_0$ vanishes. For later convenience we define the dimensionless
tuning parameter $x\equiv \g_p/\g_l-1$.

Plugging the amplitude-phase representation of $\psi$ into Eq.~(\ref{EOM}), and linearizing in the amplitude fluctuations $\chi$,  we obtain the pair of equations
\begin{widetext}
\bea
\partial_t\chi&=& -2 u_d M_0^2 \chi-K^x_c M_0\partial^2_x\t-K^y_c M_0\partial^2_y\t- K^x_d M_0(\partial_x\t)^2-K^y_d M_0(\partial_y\t)^2+\text{Re}\zeta ~,\label{chiEOM}\\
M_0\partial_t\t&=&-2 u_c M_0^2\chi+K^x_d M_0\partial^2_x\t+K^y_d M_0\partial^2_y\t - K^x_c M_0(\partial_x\t)^2-K^y_c M_0(\partial_y\t)^2+\text{Im}\zeta ~,
\label{thetaEOM}
\eea
\end{widetext}
where we have used the freedom discussed earlier to choose $r_c=-u_c M_0^2$  to simplify this expression.

Note that if we have no dissipation ($H_d=0$), so that $u_d=0$, both $\chi$ and $\t$ are ``slow'' variables, in the sense of evolving at rates that vanish as the wavevector goes to zero. In this case   we can substitute Eq.~(\ref{chiEOM})  into the time derivative of Eq.~(\ref{thetaEOM}) to obtain a wave equation for $\t$ supplemented by irrelevant non-linear corrections. This gives the linear dispersion of the undamped Goldstone modes characteristic of a lossless condensate with exact particle number conservation. In contrast, without particle number conservation  (i.e., in the presence of loss and drive), $u_d\ne 0$, and we can therefore neglect the $\partial_t\chi$ term  (which vanishes as frequency $\omega\rightarrow 0$)  on the left hand side of Eq.~(\ref{chiEOM}) relative to the $2 u_d M_0^2 \chi$ on the right hand side for any ``hydrodynamic mode'' (i.e., in the low frequency limit). Doing so turns Eq.~(\ref{chiEOM}) into a simple linear algebraic equation relating $\chi$ to spatial derivatives of $\t$. Substituting the solution for $\chi$  of this equation into Eq.~(\ref{thetaEOM}) gives Eq.~\eqref{akpz}, a closed equation for $\t$. The noise variable in that equation is related to the original noise through ${\bar \zeta}=\left(\text{Im}\zeta-u_c\text{Re}\zeta/u_d\right)/M_0$, and hence $\av{{\bar\zeta}(\vx,t){\bar\zeta}(\vx',t')}=2\Delta\d^d(\vx-\vx')\d(t-t')$ with $\Delta$ given in Eq.~\eqref{kpznoise}. The stochastic equation for $\theta$ includes all  terms that are marginal and relevant  by canonical power-counting, while neglecting irrelevant terms like $\partial_t^2\t$, $\partial_t\nabla^2\t$, and $\partial _t(\nabla\t)^2$.

\begin{acknowledgements}
Illuminating discussions with Daniel Podolsky and Anatoli Polkovnikov are gratefully acknowledged. We thank Austen Lamacraft for insightful comments on the manuscript. JT thanks Dung-hai Lee and the Physics Department of the University of California, Berkeley, CA, and the MPIPKS, Dresden, Germany, for their support (financial and otherwise) and hospitality while this work was being done.   EA likewise thanks the Miller Institute at UC Berkeley and the Aspen Center for Physics under NSF Grant \# 1066293. LC  also thanks the MPIPKS for support and hospitality, and acknowledges support by the National Science Foundation of China
(under Grant No. 11004241). We also thank the  US NSF for support by
awards \# EF-1137815 and 1006171; and the Simons Foundation for support by award \#225579 (JT), ERC grant UQUAM, and the ISF (EA), the National Science Foundation of China
(Grant No. 11004241) (LT), and the Austrian Science Fund (FWF) through the START Grant No. Y 581-N16 and the SFB FoQuS (FWF Project No. F4006- N16) (LMS, SD).
\end{acknowledgements}

\clearpage
\appendix

\section{Supplementary Material}
\subsection{Polariton condensate model with reservoir}
In the main text we worked with a Ginzburg-Landau model including only the lower polariton band. Such a model clearly gives the correct universal physics. However, in order to find how the parameters of the anisotropic KPZ equation change as actual experimental parameters are varied requires to start from a more microscopic model of the polariton degrees of freedom. 

The standard model for describing these systems is a ``two fluid'' model which includes the particles in the ``upper-polariton band'' acting as a reservoir with local density $n_R$ for the condensate which forms in the ``lower polariton band''~\cite{Carusotto2012}. Here we generalize the model slightly in order to include dissipative mass terms and anisotropy:
\begin{widetext}
\bea
\partial_t\psi&=&\left[\sum_{\a=x,y} (iK_c^\a+K_d^\a) \partial_\a^2-ir_c-\gamma_l-iu_c |\psi|^2+ R n_R\right]\psi+\zeta,\nn\\
\partial_t n_R&=&P-R n_R |\psi|^2-\g_R n_R ,
\eea
\end{widetext}
where $\av{\zeta^*(\mathbf{x},t)\zeta(\mathbf{x}',t')}=2\sigma \d(\mathbf{x}-\mathbf{x}')\d(t-t')$.
It is usually assumed that the reservoir relaxation time $\g_R$ is faster than all other scales.
Hence we may solve the reservoir density independently assuming it is time independent
\be
n_R= {P\over \g_R+ R|\psi|^2}.
\ee
Substituting this in the equation for $\psi$ we obtain
\begin{widetext}
\be
\partial_t\psi=\left[\sum_\a(iK_c^\a+K_d^\a)\partial_\a^2-i r_c-\gamma_l-i u_c |\psi|^2+{ P \over \eta+|\psi|^2}\right]\psi+\zeta ,
\ee
\end{widetext}
where we have eliminated $R$ and $\g_R$ for the single parameter $\eta=\g_R/R$. We note that the amplitude of the white noise is given by the total loss rate ($\g_l$) and gain, and since in steady state the loss and gain must be equal we simply have $\s=\g_l$  in this case.  

In the following, as in the main text we work in the phase-amplitude
representation $\psi(\vx,t)=(M_0+\chi(\vx,t))e^{i\t(\vx,t)}$ and expand around
the homogeneous mean field solution. Let us therefore first solve for the mean
field steady state. The real part of the equation gives $\g_l=P/(\h+ M_0^2)$
from which we can deduce the condensate density $M_0^2=P/\g_l-\h$. The imaginary
part of the equation is $r_c= -u_c M_0^2$. It is also worth noting that loss
comes only from the term $\g_l$, since there is no two-particle loss term in
this model (instead saturation is reached due to the non-linear reduction of the
pump term). Hence in steady state, when loss is equal to gain, the noise term is
simply $\s=\g_l$.

We now proceed to write the equations of motion for $\chi$ and $\theta$ to linear order in $\chi$. This gives
\begin{widetext}
\bea
M_0^{-1}\partial_t\chi&=&-2\g_l^2 P^{-1}M_0 \chi-K_c^\a \partial_\a^2\theta-K^\a_d(\partial_a\t)^2 +M_0^{-1}\text{Re}\zeta,\nn\\
\partial_t\t&=&-2u_c\chi+ K^\a_d\partial^2_\a\t- K^\a_c(\partial_\a\t)^2 -M_0^{-1}\text{Im}\zeta.
\eea
\end{widetext}
Now as in the main text we can eliminate $\chi$ to obtain the KPZ equation for $\theta$,
where $\alpha=x,y$ is summed over and 
\be
\partial_t\t=D^\a\partial_\a^2\t+\half\lambda^\a(\partial_\a\t)^2+\bar{\zeta},
\ee
where
\be
\bar{\zeta}=M_0^{-1}\left(\text{Re}\zeta-{u_cP\over\g_l^2}\text{Im}\zeta\right).
\ee
The noise parameter  in $\av{\bar{\zeta}^*(\mathbf{x},t)\bar{\zeta}(\mathbf{x}',t')}=2\D\d(\mathbf{x}-\mathbf{x}')\d(t-t')$  
is here given by:
\bea
\D&=&{\g_l^2/2\over P-\h\g_l}\left(1+ {u_c^2P^2\over\g_l^4}\right)
=\left(1+{u_c^2\h^2\g_p^2\over \g_l^4}\right){\g_l^2/2\h\over \g_p-\g_l}\nn\\
&=& {u_c\bar{\g}\over 2x} \left(1+{(1+x)^2\over {\bar\g}^2}\right),
\label{Delta}
\eea
where we have defined $\g_p\equiv P/\eta$, the dimensionless tuning parameter $x=\g_p/\g_l-1$ and the dimensionless loss parameter $\bar{\g}\equiv \g_l/(\eta u_c)$. Below we will also need the dimensionless interaction strength $\bar{u}\equiv u_c/\sqrt{K_c^x K_c^y}$ and the ratios $\nu_\a=K^\a_d/K^\a_c$.
The parameters of the anisotropic KPZ equation may now be written as: 
\bea\label{D-lambda}
D_\a&=& K_c^\a\left({K_d^\a\over K_c^\a} +{u_c P\over \g_l^2}\right)=K_c^\a\left(\nu^\a+{1+x\over \bar{\g}}\right),\\
\lambda_\a&=&2K^\a_c\left({K^\a_d u_cP\over K^\a_c \g_l^2} -1\right)=2K^\a_c\left(\nu^\a{1+x\over {\bar\g}}-1\right).\nn
\eea

In order to make contact to the main text, we note that the expressions for the diffusion constants $D_\alpha$ and non-linear coefficients  $\lambda_\alpha$ can be obtained from   the predictions Eq.~(\ref{D-lambda}) for the Ginzburg-Landau model Eq.~(\ref{EOM}), if we make the replacement  
\be\label{eq:rep}
u_d={\g_l^2\over P}={u_c \bar{\g}\over 1+x}.
\ee
The parameter $K_d$ is thought to be small in isotropic two-dimensional quantum wells. If we take $K_d=0$, then $D=K_c u_c/u_d=K_c(1+x)/{\bar\g}$, and $\lambda=-2K_c$. 
Eq.~\eqref{g_02band} in the main text  corresponds to this special case $K_d=0$.

\subsection{Crossover scales in isotropic polariton condensates}

We will now use the results just presented for  
the isotropic case without dissipative mass terms; i.e., $\nu_x=\nu_y=0$ (for the anisotropic case, see main text). This implies $D=K_c(1+x)/2{\bar \g}$ and $\lambda=-K_c$. The dimensionless coupling constant $g$  is then given by
\bea
Ög&=&{\D\lambda^2\over D^3}={2\bar{u} \bar{\g}^2}\,\left({\bar{\g}^2+(1+x)^2 \over x (1+x)^3}\right).
\eea
From the expression for $g$ we can extract the dependence of the KPZ length on the tuning parameter:
\bea
\log(L_*/\xi_0)=\left({4\pi\over \bar{u}\bar{\g}^2}\right){x(1+x)^3\over \bar{\g}^2+(1+x)^2}.
\eea

\subsection{Critical locus in the anisotropic case.}

The critical point for vortex unbinding can be estimated by solving for the renormalized  scaled noise $\kappa(\ell\to\infty)\equiv\kappa(\infty)$ as a function of the bare value using the RG equations of the non-compact KPZ equation; this involves additional recursion relations for $D_\alpha$ and $\D$ as well as Eq.~(\ref{recrels}); for details see reference~\cite{Chen2013}. This  analysis gives $\kappa(\infty)=-\kappa_0(1-\G_0)^2/(4\G_0)$. The KT transition occurs at the point where this renormalized value $\kappa(\infty)$ of $\kappa$ reaches $\pi$. Hence 
the  phase boundary in the $\kappa_0$-$\G_0$ plane is then a locus %given by the curve
in the plane of bare scaled noise and anisotropy parameters given by ~\cite{Chen2013}
\begin{eqnarray}
\kappa_0=-{4\pi\Gamma_0\over\left(1-\Gamma_0\right)^2}~.
\label{phaseboundary}
\end{eqnarray}

\end{document}